# The Design Principles of Konrad Zuse's Mechanical Computers

Raul Rojas
Freie Universität Berlin
December 2015

Abstract

*Konrad Zuse built the Z1, a mechanical programmable computing machine, between 1935/36 and 1937/38. The Z1 was a binary floating-point computing device. The individual logical gates were constructed using metallic plates and interconnection rods. This paper describes the design principles Zuse followed in order to complete a complex calculating machine, as the Z1 was. Zuse called his basic switching elements "mechanical relays" in analogy to the electrical relays used in telephony.*

## 1    Introduction

The German inventor Konrad Zuse built his first computing machine in the apartment of his parents (who unfortunately did not have a silicon age style garage) during the years 1935 to 1937 [1]. The machine, called afterwards the Z1 by Zuse himself, has been hailed as the first programmable calculator in the world. The Z1 was well ahead of its time. Although the Harvard Mark I and UPenn's ENIAC were built several years after the Z1, they were not fully binary computers, as the Z1 was [2]. Moreover, the Z1 used floating-point binary arithmetic for computing the four basic arithmetical operations. The program was punched in a tape using a binary code. Numbers could be stored in memory or could be retrieved from it. Only the conditional jump was missing in the instruction set. Otherwise, the Z1 would have been a universal computing device, although tortuous programming tricks can be used to obtain universality from machines able to process a single loop of arithmetical operations.

Recently, I described the main dataflow architecture of the Z1 [3]. In his old years, Konrad Zuse built a reconstruction of the Z1 (delivered to a museum in 1989). Using the blueprints of that reconstruction as a reference, as well as other historical descriptions [4], it was possible to decode the computer architecture of the Z1. Such archeological endeavor was necessary because the Z1 was a mechanical device based on outdated technology. The Z1 reconstruction, on display in Berlin's Technology Museum, looks like a clever set of plates, rods and bars, a sort of intricate clockwork almost impossible to make sense of (nonetheless, we have implemented a simulation of part of the logic [5]). In this paper, I provide a guide to the main design principles used by Konrad Zuse in his mechanical devices. The Z1 was not the last mechanical circuit built by Zuse: the memory of the Z4 computer, finished in 1945, was made of mechanical components but the logic of the Z4 was constructed using telephone relays. Until late in the 1940s, Zuse was still hoping that advances in mechanical manufacturing would make mechanical memories competitive with memories made of vacuum tubes. The high cost of the latter would offset the low speed of the mechanical components – so Zuse thought. The invention of the transistor put an end to such speculations.

## 2     The mechanical relay

The basic logical component used by Konrad Zuse in the Z1 was the "mechanical relay". All gates were binary. The logical components could move only one step in one direction (Zuse arranged the components on a plane so that the allowed directions were West, South, East, and North). The initial state of any component was the state 0. Its state after a linear movement was state 1. The components could move back and forth between state 0 and 1.

Fig. 1 shows a diagram of a mechanical relay. Bit A is called the "control bit" or "control element". On the left side of Fig. 1 we see the case where the initial state of bit A is such that there is no mechanical connection between the actuator and the actuated plate. The movement of plate B is not copied to bit C, if A is zero. However, if A moves down to its position 1, then mechanical coupling is achieved and the movement of plate B is copied to bit C. This mechanical relay is then just a one-step delay.

The right of diagram 1 shows the case where the initial state of A (that is A=0) is such that the mechanical coupling between plates B and C is present. If A moves to state 1 (up) then the moving plate B loses its mechanical coupling to plate C. In this case bit C will be the negation of bit A: when bit A is 0, C is 1 at the clock signal. When bit A is 1, bit C is 0.

Fig. 2 shows how to implement the logical gates AND and OR using two mechanical relays. In the case of the AND circuit, the mechanical movement of the actuator plate (activated when a clock signal arrives) is only transmitted when A=B=1. In the case of the OR circuit, movement is transmitted when A or B are equal to 1.

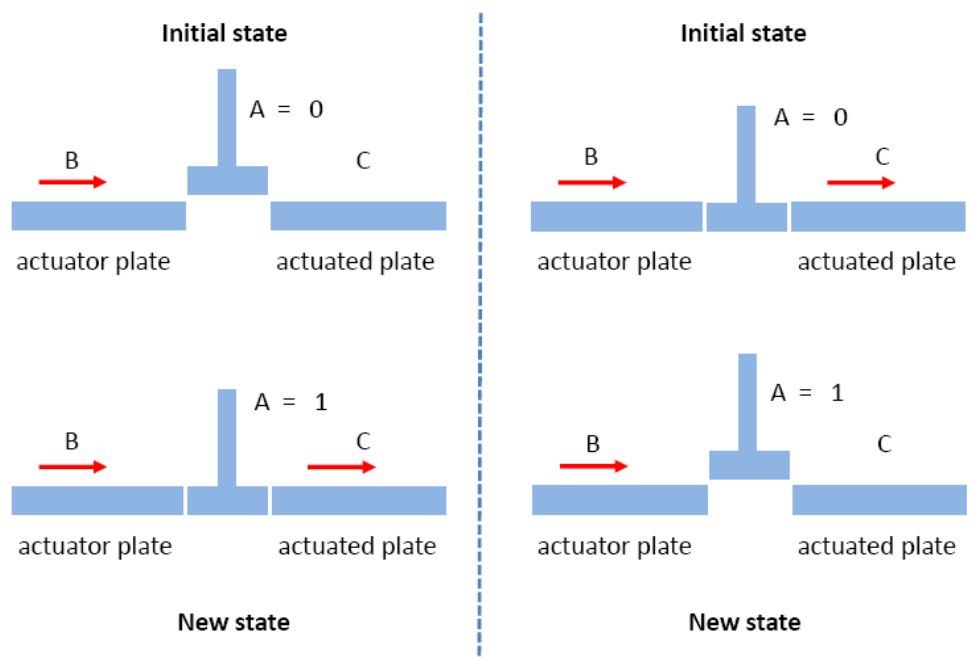

Fig. 1: A mechanical relay with two different initial states of the control plate. On the left, the initial state does not provide mechanical coupling. On the right, the initial state provides mechanical coupling. The initial state is called 0, the state after a displacement is called a 1.

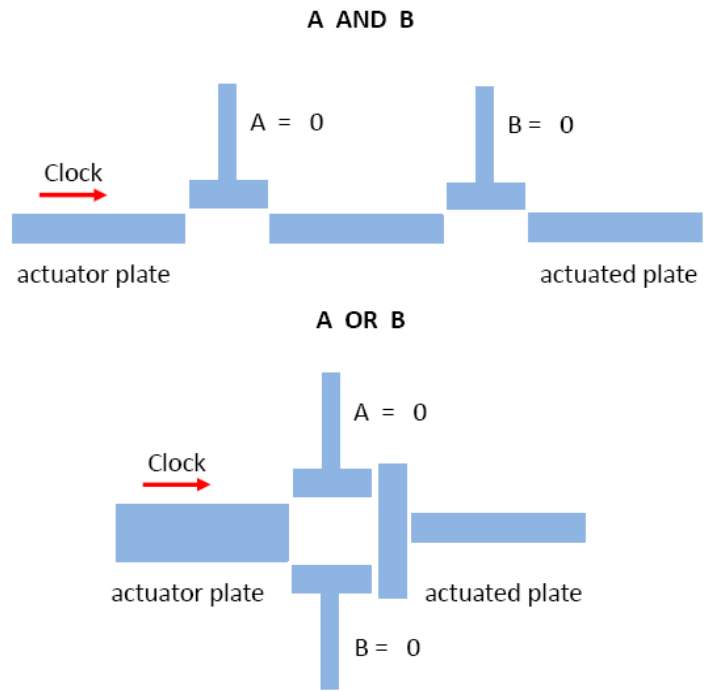

Fig. 2: Two mechanical logical gates: AND on top, OR at the bottom.

One interesting aspect of such mechanical constructions is that a long logical formula can be computed with zero delay, in principle, by such mechanical arrangements (if we assume that movement between rigid plates is transmitted instantaneously). A conjunction of 100 bits, for example, could be computed by the concatenation of 100 mechanical relays. Once all 100 control bits have been set, the actuator plate will move the actuated plate only when all bits are equal to 1.

It is easy to see that any kind of logical gate can be constructed using a mechanical relay. An XOR, for example, can be obtained as a variation of the AND gate with different initial conditions for the bits A and B, as shown in Fig. 3. Bit A moves down, Bit B moves up.

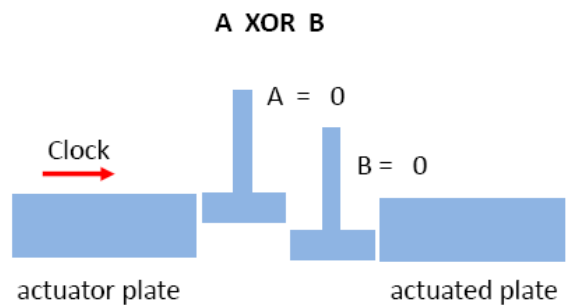

Fig. 3: An XOR gate made of mechanical relays.

Now, Zuse did not use exactly this kind of mechanical arrangements, because the main problem is making sure that all parts will move and will come back to their initial position. Control plates (such as bit A in Fig 1), move in one direction and then in an orthogonal direction when the movement of the actuator plate is transmitted. It is not easy to do this with mechanical components. Therefore, Zuse's idea was to use small vertical rods as "connectors" and to let them move between two horizontal planes made of metal or glass. Just think of plate A and B in Fig. 3 as connecting through a single rod. Keeping everything together, however, requires many different plates, sometimes duplicated. This is what produces the obscure mechanical drawings completed by Zuse during the different stages of the construction of the Z1. The main idea, as we saw before, is simple; only its mechanical realization is somewhat involved. Fig. 4 shows an example of the kind of metallic constructions used by Zuse. Here the "data bit" corresponds to the control plate, and it can move between the zero and one position. When the data bit is in state 1, the rod crossing all three plates can transmit the movement of the actor plate (which in this case is pulled) to the actuated plate. When the data bit is zero (left side of Fig. 5) the movement of the actuator plate is not transmitted.

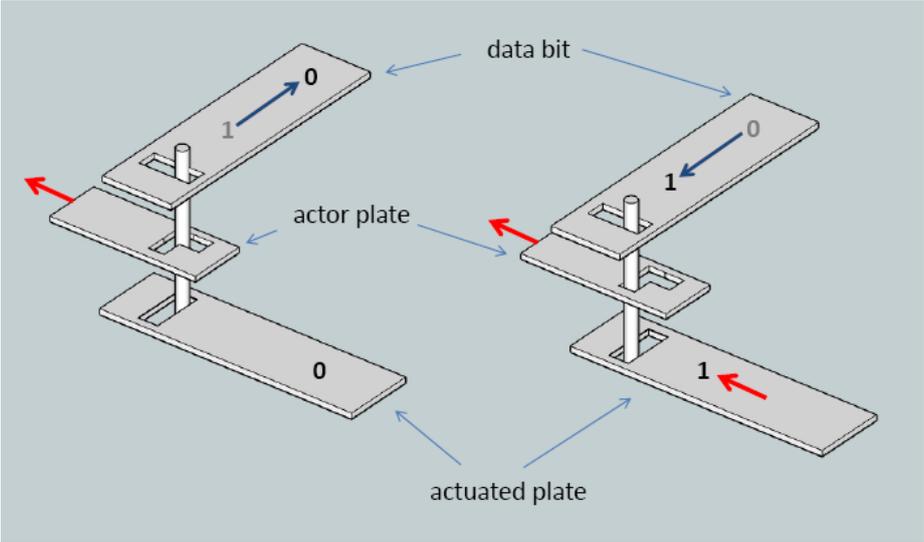

Fig. 4: A mechanical relay of the type used by Konrad Zuse

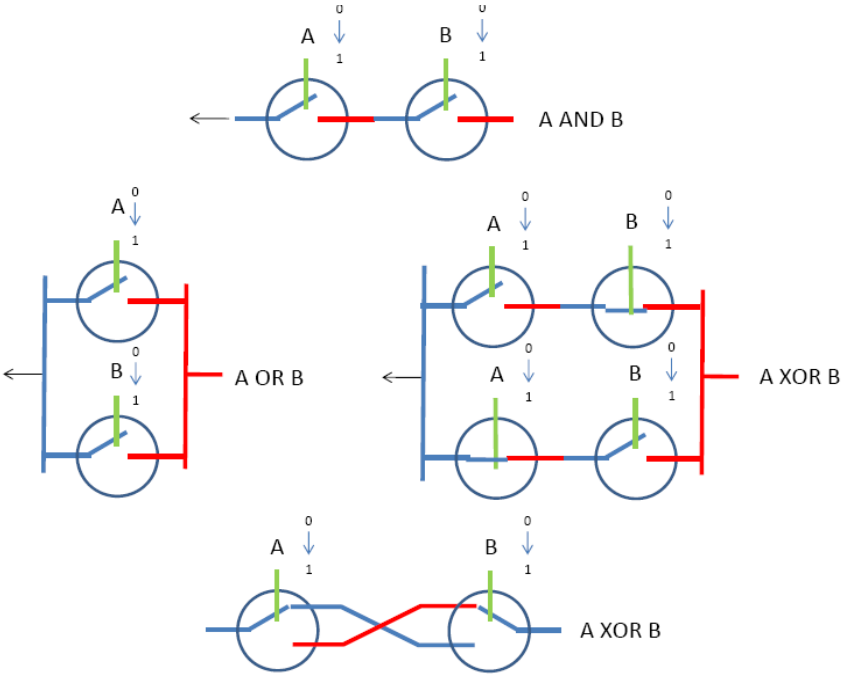

Fig. 5: Zuse's notation for mechanical relays and some elementary gates.

Zuse developed an abstract notation for his relays which abstracts from the mechanical nature of the devices and stresses the logical properties. Fig. 5 is a summary of some logical gates constructed out of elementary mechanical relays. As always, the initial position shown in the diagram corresponds to the zero state. Movement corresponds to the one state. A relay can opened or closed when it switches from state zero to state one.

Zuse developed mechanical gates where the movement could be made by pushing with the actuator plate (as in my examples in this section), but it is also possible to design variations where the actuator plate can pull the actuated plate (Fig. 4). In Fig. 5 this possibility is revealed by the direction of the arrows.

## 3    The mechanical clock cycle

A complex computer requires the presence of feedback loops in its circuits. The result of a previous step must be fed back to the processor for further computation. In order to reduce the degrees of freedom during the design process, Zuse settled for the four directions of movement mentioned before (W,S,E,N) and introduced the "common cycle", that is a clock cycle subdivided in four subcycles. During subcycle I a pulse in the West direction, for example, was transmitted from the clock unit to the machine. During subcycles II, III, and IV, the movements transmitted went in the South, East, and North directions respectively. In a mechanical relay, the control plate could then be activated in subcycle I, the actuator plate transmitted its movement in subcycle II. In subcycle III the control plate retracted to its original position and in subcycle IV, the actuator and actuated plates could also retract to their original positions.

The clock cycle in the Z1 was provided by a crank, which could be rotated manually or with the help of an electric motor (Fig. 7, Fig. 8). This means that the effective cycle time of the Z1 was totally "user dependent", it could be any value below the maximum allowable speed so that the components would not be subjected to excessive mechanical stress. The typical cycle time of the Z1, quoted in some papers, is 5 Hz.

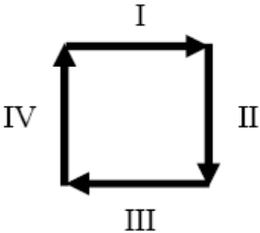

Fig. 6: Zuse's "Einheitskreislauf" (common cycle)

In a complex circuit the actuated plate could act as the control plate for another circuit. If the control plate for the first circuit was set at subcycle I, the first circuit could be actuated at subcycle II and the result could be used as the control bits for a subcircuit activated at subcycle III, and so on. With this arrangement the maximum depth of a circuit which started its computation and finalized within a clock cycle was depth 3. Fortunately, the maximum depth of a circuit for computing the addition of binary numbers is precisely three.

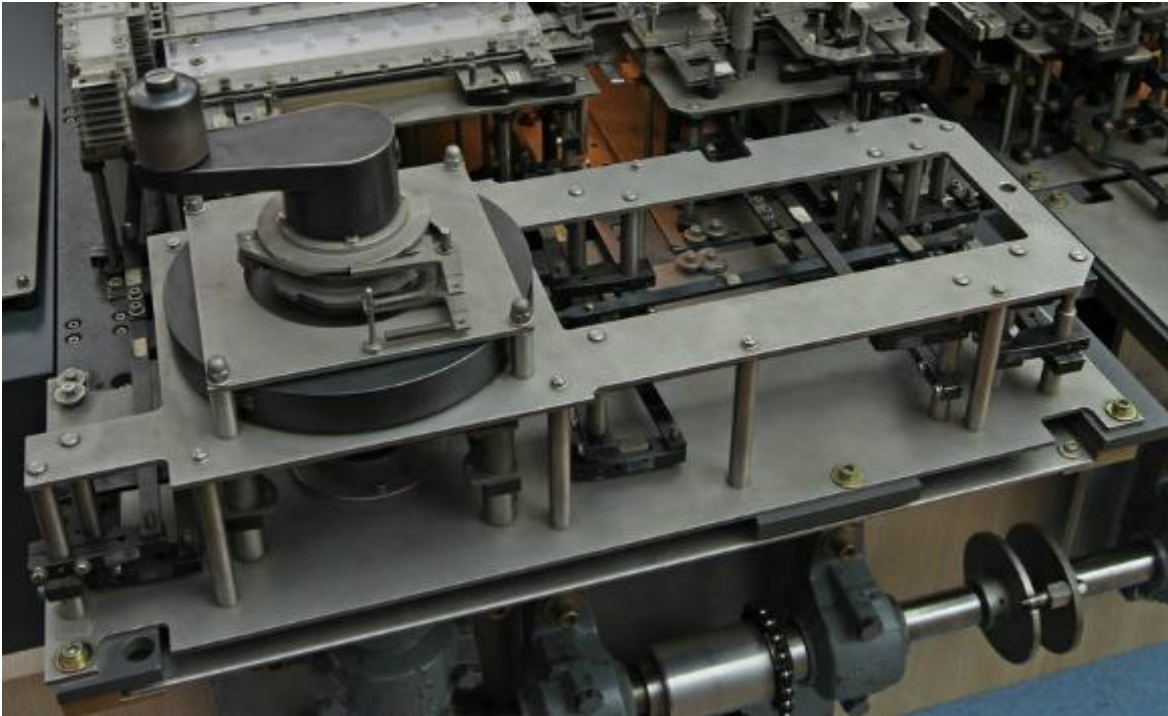

Fig. 7: The crank for producing the common cycle. Notice the levers used for transmitting the four directions of movement all across the machine.

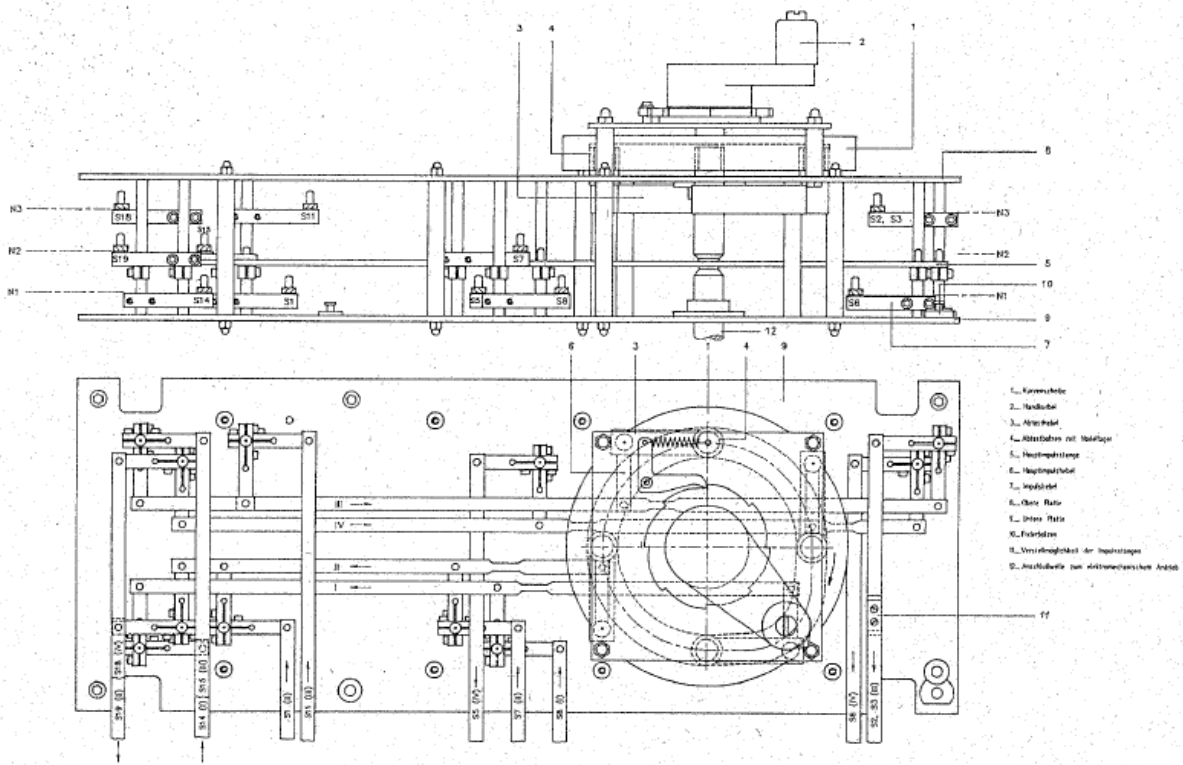

Fig. 8: The diagram of the common cycle mechanism [6]. The four subcycles are sent in one parallel direction but can be turned 90 degrees using levers.

## 4     Transmission of impulses

Zuse built the reconstructed Z1 in such a way that all the clock subcycles are transmitted using rods and levers located in the "basement" of the machine. The logic is located in the upper part of the Z1, distributed across several layers, like a logical sandwich.

Whenever the state bits are represented by movement across four possible directions, it is necessary to provide means of changing the direction of movement. This can be done using levers, as shown in Fig. 9.

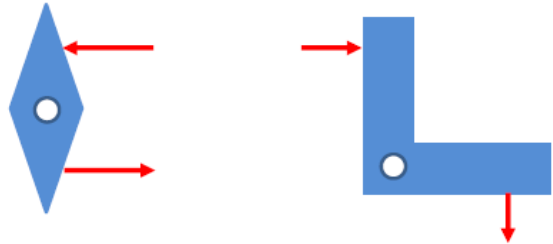

Fig. 9: Mechanical movement: direction changers

It is also possible to have something like an electrical "rectifier" that only allows logical movement to flow in one direction. This is done by using mechanical gates where the actuator plates only push the actuated plate or plates, without using a rod going through several plates (as is the case in Fig. 4). This would be more in the spirit of the OR gate shown in Fig. 2, where any of the control plates A or B, or both, can push the actuated plate. Zuse had a special notation (using an arrow) for such independently "pushed" plates. He called this "rectification", like in electrical circuits when electricity can only flow in one direction.

One important problem Zuse had to deal with, is the case when a circuit uses its result as new data. Such is the case of the arithmetical unit (ALU), where a partial result has to be fed back to the ALU for further processing, for example, during a multiplication performed by repeated addition. But here we have a contradiction: when one circuit finishes its calculation it has to be brought back to its initial state (by the movements allowed by the common cycle). Partial results have to be captured and have to be delayed until they are needed again in the ALU. In order to handle this, Zuse designed a "delay line", which is nothing but a sequence of mechanical relays, positioned one after the other. The result of a relay is used as the control bit for the next relay, and so on. The actuator plates are activated by successive movement subcycles. Any number of subcycles can thus be interposed between the production of a result and its subsequent use, without having to use a memory cell for holding this value.

## 5      A worked-out example: the mechanical addition unit

Fig. 10 shows the basic building block for a mechanical adder (as proposed by Zuse in [4]). The figure shows the addition of the *i*-th column in the sequence of bits $a_k a_{k-1}...a_0$ to be added to the sequence of bits $b_k b_{k-1}...b_0$. The bit $a_i$ is set in the subcycle previous to subcycle I. In this example, subcycle I moves in the North direction and the subsequent subcycles rotate their direction of movement by 90 degrees. The bit $b_i$ is pulled up (if it is a one) at the same time as subcycle I becomes active. At the end of subcycle I the first two gates from the top have computed the conjunction and disjunction, respectively, of the two bits. The carry-bit from the previous column of bits ($c_{in}$) is pulled to the right at subcycle II. A carry-bit for the next column of bits ($c_{out}$) is generated if the two input bits were (1,1), or if they were one of the pairs (0,1) or (1,0), and there is an active carry from the previous column ($c_{in}$).

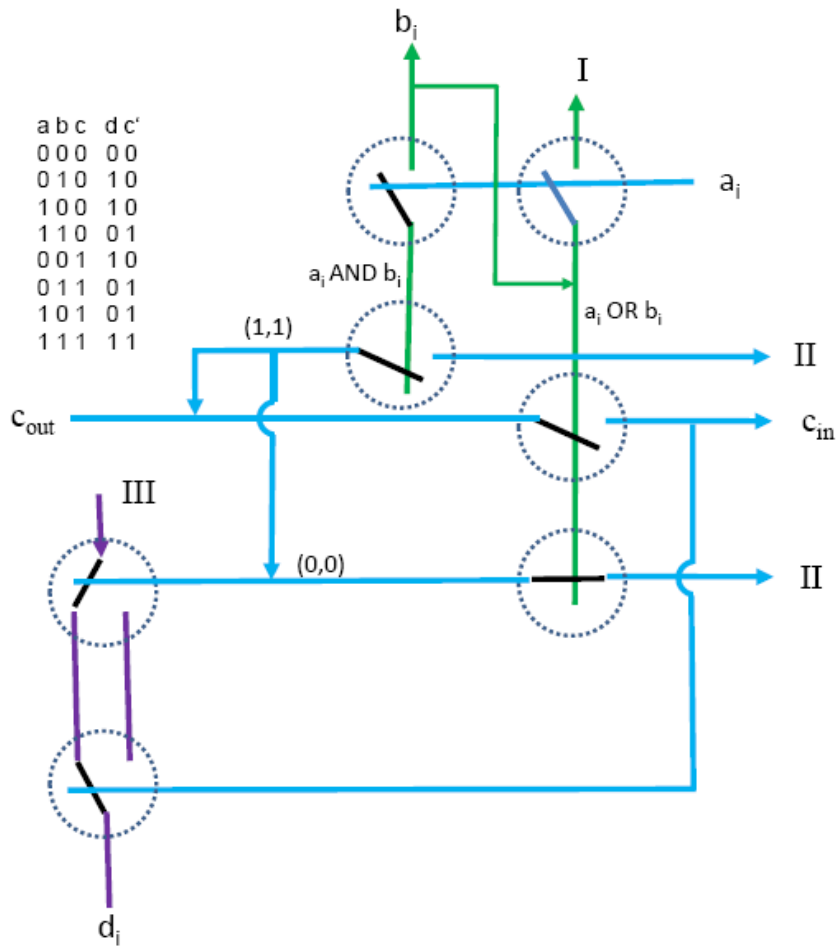

Fig. 10: Circuit for the addition of two bits $a_i$ and $b_i$ with a carry $c_{in}$, and generation of the next carry bit $c_{out}$.

The circuit of two gates activated by subcycle III computes the addition bit. If the input was (1,1) or (0,0), the carry-bit from the previous column has to be one, in order to have $d_i$=1. If the input was (0,1) or (1,0), the carry-bit from the previous column has to be zero in order to have $d_i$ =1. I all other cases $d_i$ is equal to zero.

## 6   Summary

This paper has presented the basic design principles followed by Konrad Zuse in his mechanical designs. They can be summarized as follows:

- The state zero of a bit corresponds to non-movement from the initial position.
- The state one of a bit corresponds to a movement step along one of four possible orthogonal directions.
- In a mechanical relay, the actuator plate can either push or pull the actuated plate.
- Several actuated plates can push on the same actuated plate (rectification).
- The actuated plate of a mechanical relay can be the control plate of another mechanical relay.
- Mechanical relays can be combined to produce all logical gates.

- In order to simplify the circuits, a "common cycle" consisting of four orthogonal movements, named I, II, III, and IV is used. The subcycles allow the machine to synchronize its computations.
- If a mechanical relay activates the control plate at subcycle I, II, III, or IV, it returns the control plate to the original position at subcycle III, IV, I, or II, respectively.
- If the actuated plate is moved at subcycle I, II, III, or IV, it returns to its original position at subcycle III, IV, I, or II, respectively.
- Movement along any of the four orthogonal directions can be reversed or can be rotated by 90 degrees, clockwise or counterclockwise, using levers.
- A mechanical relay can be used as a delay element (since a mechanical relay just copies a bit after one subcycle).

This having been said, it must be pointed out that the vertical rods used by Konrad Zuse in most of his mechanical relays (Fig. 4 and Fig. 11) were an accident waiting to happen. The rods moved vertically, sandwiched between glass or metallic plates, and had to be pulled or pushed symmetrically and gently, so that they will not topple over. This was achieved by duplicating the actuated and actuator plates. The result was never completely satisfactory, except for the mechanical memories built by Zuse, which were still being used for the Z4 in the 1950s.

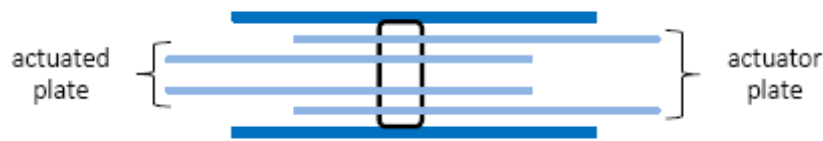

Fig. 11: A mechanical relay and its control rod sandwiched between plates.

Based on the principles listed above, Zuse's mechanical diagrams read very much like electrical circuits and can be used to build machines based on telephone relays. This is exactly what Zuse did after it was obvious that the mechanical stress in his mechanical constructions was excessive. After the Z1, Zuse built the Z3 using telephone-relays. It was shown in 1941 computing small determinants.